\documentclass{SCGE}
\usepackage[colorlinks,linkcolor=blue,citecolor=blue,urlcolor=blue]{hyperref}

\let\citedash\relax
\makeatletter \providecommand{\citedash}{\hbox{-}\penalty\@m}
\makeatother

\usepackage{multirow, array}

\begin{document}
\bibliographystyle{unsrt}  
\begin{picture}(0,0){\rm
\put(0,-20){\makebox[160truemm][l]{\bf {\sanhao\raisebox{2pt}{.}}
Article  {\sanhao\raisebox{1.5pt}{.}}}}}
\put(0,-34){\jiuwuhao {\textcolor[rgb]{0.5,0.5,0.5}{\sf 
}}}
\end{picture}

\def\bm{\boldsymbol}

\def\dl{\displaystyle}
\def\du{\end{document}}

\def\be{\begin{eqnarray}}
\def\ee{\end{eqnarray}}
\def\bh{{\rm \bullet}}
\def\sph{{\rm sph}}
\def\d{{\rm d}}
\def\x{{\rm X}}
\def\agn{{\rm AGN}}
\def\gal{{\rm Gal}}
\def\bol{{\rm bol}}
\def\edd{{\rm Edd}}
\def\acc{{\rm acc}}
\def\s{{\rm s}}
\def\r{{\rm R}}
\def\bh{\bullet}

\Year{2017} %
\Month{October} %
\Vol{60} 
\No{10} 
\BeginPage{1} 
\AuthorMark{{\rm X. Zhang}, et al.}  
\DOI{} 
\ArtNo{109511}

\title[ON THE MEAN RADIATIVE EFFICIENCY OF ACCRETING MASSIVE BLACK
HOLES IN AGNs and QSOs]{On the mean radiative efficiency of accreting 
massive black holes in AGNs and QSOs}

\author[1, 2*]{Xiaoxia Zhang}{}
\footnote{*Corresponding author (Youjun Lu, email: luyj@nao.cas.cn; \\
Xiaoxia Zhang, email: zhangxiaoxia16@gmail.com)}
\author[1, 2*]{Youjun Lu}{}

\address[{\rm1}]{National Astronomical Observatories, Chinese Academy of
Sciences, Beijing 100012, China}
\address[{\rm2}]{School of Astronomy and Space Sciences, University of Chinese
Academy of Sciences, No. 19A Yuquan Road, Beijing 100049, China}

\maketitle \vspace{-3.5mm}{\footnotesize\begin{center} 
\end{center}}\vspace*{-5mm}

\begin{center}
\rule{16.5cm}{0.4pt}
\parbox{16.5cm}
{\begin{abstract} 

Radiative efficiency is an important physical parameter that describes
the fraction of accretion material converted to radiative energy for
accretion onto massive black holes (MBHs). With the simplest So{\l}tan
argument, the radiative efficiency of MBHs can be estimated by
matching the mass density of MBHs in the local universe to the
accreted mass density by MBHs during AGN/QSO phases. In this paper, we
estimate the local MBH mass density through a combination of various
determinations of the correlations between the masses of MBHs and the
properties of MBH host galaxies, with the distribution functions of
those galaxy properties. We also estimate the total energy density
radiated by AGNs and QSOs by using various AGN/QSO X-ray luminosity
functions in the literature. We then obtain several hundred estimates
of the mean radiative efficiency of AGNs/QSOs. Under the assumption
that those estimates are independent of each other and free of
systematic effects, we apply the median statistics as described by
Gott et al.\cite{got01} and find the mean radiative efficiency of 
AGNs/QSOs is $\epsilon=0.105^{+0.006}_{-0.008}$, which is consistent 
with the canonical value $\sim0.1$. Considering that about $20\%$ 
Compton-thick objects may be missed from current available X-ray 
surveys, the true mean radiative efficiency may be actually $\sim 0.12$.
\end{abstract}}
\end{center}\vspace*{-0.6cm}

\begin{center}
\parbox{16.5cm}
{\bf\jiuhao black hole, galaxies, quasars }
\end{center}

\begin{center}
{}
\CITA    
\end{center}

\textwidth=178truemm \textheight=236truemm

\wuhao\vspace*{1.5mm}

\begin{multicols}{2}

\renewcommand{\baselinestretch}{1.08} \baselineskip 12.2pt\parindent=10.8pt

\renewcommand{\thefootnote}

\section{INTRODUCTION}

Radiative efficiency ($\epsilon$) is an important physical parameter
that describes the fraction of accretion material converted to
radiative energy for accretion processes. For disk accretion on to a
black hole (BH), the value of $\epsilon$ is directly determined by the
spin of the BH, as well as the rate and geometry of the accretion flow [e.g.,
thin disk accretion, thick disk accretion, or advection dominated
accretion flow (ADAF)]. Most QSOs may accrete via thin disk accretion
and are expected to radiate at an efficiency in the range from $0.04$
for retrograde disk accretion to $0.42$ for prograde disk accretion
onto maximally spinning massive black holes (MBHs), an order of
magnitude difference in $\epsilon$ \cite{LyndenBell69,
Bardeen70}.  Accurately estimating radiative efficiencies of QSOs is
of great importance as it may help to reveal the spins and also the
evolution of their central MBHs, the mode and detailed physics of the
accretion flow onto the MBHs.

There are a few methods to estimate the radiative efficiency
$\epsilon$ of QSOs in the literature. These methods include: (1)
direct multi-wavelength continuum fitting by adopting thin disk
accretion model \cite{Sun89}; (2) combination of direct
bolometric luminosity estimation from multi-wavelength spectra and
accretion rate estimation from optical continuum fitting for those
QSOs with MBH mass measurements \cite{Davis11, wu13}; (3) global
constraint from connecting the local MBHs with distant AGNs/QSOs via
the So{\l}tan argument and its variants
\cite{sol82, yu02, Elvis02, mar04, Shankar09}. Methods (1)
and (2) can provide direct estimations of $\epsilon$ for many
individual AGNs/QSOs, but those estimations suffer from significant
uncertainties, not only because they are model dependent, but also
because complete multi-wavelength spectra from radio to X-ray are not
available for many AGNs/QSOs.  The global constraint on the radiative
efficiency $\epsilon$, obtained from the (extended) So{\l}tan
argument, method (3), may be more accurate as it does not depend on
the detailed accretion physics, though it only gives a ``mean'' value
of $\epsilon$.

In the past two decades, the mean radiative efficiency of AGNs/QSOs
has been intensively estimated through the So{\l}tan
argument or its variants by connecting the local MBH mass function
with the luminosity function of distant AGNs/QSOs
\cite{yu02, Elvis02, mar04, yu04, Merloni04, cao07, yu08,
Wang09, Shankar09, Treister09, zha12, Li12, Shankar13}. Currently, the
AGN/QSO luminosity functions are quite accurately measured by many
surveys. However, it is still difficult to directly obtain the mass
function of MBHs in the local universe because the number of MBHs in
galactic centers that have mass measurements is still limited (less
than $100$).  Instead, this MBH mass function is usually estimated
from the distribution of galactic properties (either $\sigma$ or
$M_{*,\rm sph}$) by using the strong correlations between the masses
of MBHs and the properties of the MBH host galaxies. However,
different studies obtained quite different values of this ``mean''
efficiency, which cover a wide range, from $\sim 0.04$ to $0.3$. The
large uncertainties in the ``mean'' efficiency estimations are mainly
due to the following reasons: first, the normalizations and the slopes
of those correlations between MBH mass and galactic properties
determined by different groups are quite different; second, there are
some uncertainties in the estimates of the distribution of galactic
properties; and third, there are also some uncertainties in the
estimates of the AGN/QSO luminosity functions.

In this paper, we revisit the estimation of the mean radiative
efficiency of AGNs/QSOs according to the simplest version of the
So{\l}tan argument by using various determinations of the relations
between the MBH mass and their host galaxy properties, the
distribution of galaxy properties, and the AGN/QSO luminosity function
given in the literature. We also analyze the uncertainties of those
estimations.  We use the median statistics, as discussed in detail in
Gott et al. \cite{got01}, to get the best estimate of the mean radiation
efficiency. The paper is organized as follows. In
Section~\ref{sec:bhmf}, we estimate the local MBH mass density by
using various relationships between the MBH mass and galactic
properties and the distribution of galactic properties. We obtain the
total energy radiated by AGNs/QSOs by using AGN/QSO X-ray luminosity
functions in Section~\ref{sec:agn}. In Section~ \ref{sec:result}, we
consider the radiative efficiencies estimated in this paper according
to the MBH mass density in the local universe and the energy radiated
from AGNs/QSOs.  We use the median statistics to get the best estimate
about the ``mean'' radiative efficiency of AGNs/QSOs.  Conclusions are
given in Section~\ref{sec:conclusion}.

In this paper, we adopt a flat universe with $H=70 \ {\rm km\ s^{-1}\
Mpc^{-1}}$, $\Omega_{\rm M}=0.27$, and $\Omega_\Lambda=0.73$, if not
otherwise specified.

\section{Mass density of MBHs in the local universe}
\label{sec:bhmf}

The mass density of MBHs in the local universe ($\rho_{\bh,0}$) may be
estimated from the MBH mass function, i.e., 
\be
\rho_{\bh,0} =\int M_\bh \frac{\d N}{\d \log M_\bh} \d \log M_\bh,
\label{eq:mbhden}
\ee
where $\frac{\d N}{\d \log M_\bh}$ is the MBH mass function at
redshift $z=0$. Unfortunately, the MBH mass function cannot be
directly determined yet since the number of mass measurements of
dormant MBHs in nearby galaxies is quite limited ($<100$ \cite{kor13}). 
However, one may still be able to estimate the MBH
mass function according to the relations between the MBH masses and
the properties of their host galaxies.

\subsection{Relations between the masses of MBHs and the properties of
MBH host galaxies}

The masses of MBHs in galactic centers are tightly correlated with 
the properties of their host spheroids, e.g., velocity dispersions
($\sigma$), bulge luminosity ($L_{\rm bulge}$), or stellar mass
($M_{*,  \sph}$). For details, see a recent review by Kormendy and Ho 
\cite{kor13}. Denoting $X$ as one of the properties of those host 
galaxies, e.g., $\sigma$ or $M_{*,  \sph}$, the correlation between 
the MBH mass $M_\bh$ and $X$ can be described as
\be
\langle \log M_\bh \rangle= \alpha + \beta \log (X/X_*), 
\label{eq:m-x}
\ee
where $\alpha$ and $\beta$ are the normalization and the slope of the
relationship,respectively, and $X_*$ is the characteristic value of $X$. 
For the $M_\bh - M_{*, \sph}$ relation, $X_*=M_*=10^{11} M_{\odot}$; for 
the $M_\bh-\sigma$ relation, $X_*=\sigma_* = 200 \ {\rm km\,s}^{-1}$. As
suggested by observations, such a relation may also have an intrinsic
scatter of $\Delta_{\log M_\bh}$ (on the order of $\sim 0.3$~dex). 

These relations have been studied intensively over the past two
decades, and the most frequently studied ones are the $M_\bh - \sigma$
and the $M_\bh - M_{*,\sph}$ relation. Table~\ref{tab:t1}
lists the values of $\alpha$, $\beta$, and $\Delta_{\log M_\bh}$ for
these two relations determined by different authors in the literature.
As seen from Table~\ref{tab:t1}, the normalization and the slope of
each of these two relations determined by different authors are quite
different, for example, the normalization of both relations can differ
by a factor of up to 3 to 4; the slope varies from $\sim 4$ to
$\sim 5.6$ for the $M_\bh - \sigma$ relation and from $\sim0.9$ to
$\sim 1.9$ for $M_\bh - M_{*, \sph}$ relation, respectively. 

In addition, it has also been proposed that the fitting form of those
relations can be even more complicated than the simple power law form
presented in Equation~(\ref{eq:m-x}), e.g., Graham et al. \cite{gra12} 
suggested that a double power law can fit the data better than a simple 
power law, and Saglia et al. \cite{sag16} adopted a polynomial form to 
fit the data. These results are also listed in Table~\ref{tab:t1}.

\begin{tablehere}
\caption{Summary of the $M_\bh - \sigma$ and the $M_\bh - M_{*, \sph}$
relationships. The last column indicates the references in which each 
quoted relationship is obtained. The relationships given by Graham et al. 
\cite{gra12} are in a double power law form, with critical mass 
$M_{\rm c}=7\times10^{10} M_\odot$ for the $M_\bh - M_{*}$ relation and 
critical velocity dispersion $\sigma_{\rm c}=200 \ \rm {km \ s^{-1}}$ 
for the $M_\bh -\sigma$ relation, respectively. Note that the 
normalizations listed in Table~\ref{tab:t1} have been rescaled by 
adjusting the Hubble constant adopted in different papers to $H_0 = 70 \ 
{\rm km\,s^{-1}\,Mpc^{-1}}$.} \label{tab:t1}
\vspace{-1mm}\footnotesize
\begin{center} \doublerulesep 0.1pt \tabcolsep 5.0pt
\begin{tabular}{ccccc} \hline 

\multicolumn{5}{c}{\multirow{2}{*}{$\log M_\bh=\alpha+\beta \log
(X/X_*)$}} \\ \\ \hline
 %
$\alpha$  		& $\beta$		& $\Delta_{\log
M_\bh}$   		& $X$ 		& Ref.   \\
\hline
$8.15\pm0.06$		& $4.02\pm0.32$		& 0.27
&\multirow{6}{*}{$\sigma$}		& Tre02	\cite{tre02}	\\
$8.12\pm0.08$		& $4.24\pm0.41$		& 0.44		&
& Gul09	\cite{gul09}	\\
$8.32\pm0.05$		& $5.64\pm0.32$		& 0.38		&
& MM13	\cite{mcc13}	\\
$8.490\pm0.049$	& $4.377\pm0.290$		& 0.29		&
& KH13	\cite{kor13}	\\
$8.372\pm0.014$	& $4.868\pm0.32$		& 0.38		&
& Sag16	\cite{sag16}	\\
$7.8 \pm \cdots$				& $5.7 \pm \cdots$
& 0.28		&		& Sha16 \cite{sag16} \\  \hline
$8.22\pm0.10$		& $1.12\pm0.06$		& 0.3
& \multirow{4}{*}{$M_*$}  	& HR04	\cite{har04}	\\
$8.56\pm0.10$		& $1.34\pm0.15$		& 0.17		&
&MM13 \cite{mcc13} \\
$8.69\pm0.05$		& $1.16\pm0.08$		& 0.29		&
& KH13 \cite{kor13} \\
$8.580\pm0.007$		& $0.885\pm0.080$		& 0.424
& 		&Sag16 \cite{sag16} \\ \hline
$8.33\pm0.12$		& $4.57\pm1.10$		& 0.34		&
$\sigma(<\sigma_{\rm c})$	& \multirow{4}{*}{Gra12 \cite{gra12}}	\\
$8.24\pm0.14$		& $4.74\pm0.81$		& 0.28		&
$\sigma(>\sigma_{\rm c})$	&		\\
$8.68\pm0.11$		& $1.92\pm0.38$		& 0.57		&
$M_*(<M_{\rm c})$	& 	\\
$8.56\pm0.30$		& $1.01\pm0.52$		& 0.44		&
$M_*(>M_{\rm c})$ &		\\ \hline
\multicolumn{5}{c}{\multirow{2}{*}{$\log M_\bh=\alpha+\beta \log
(X/X_*)+\gamma \log^2 (X/X_*)+\delta \log^3 (X/X_*)$}} \\ &&&&\\
\hline
$\alpha$ \ \ \ \ \ $\beta$	& $\gamma$ \ \ \ \ \ $\delta$	&
$\Delta_{\log M_\bh}$   		& $X$ 		& Ref.   \\
\hline
7.574 \ \   1.946   & -0.306 \ \  -0.011 	& 0.4  	& $M_*$
& Sha16 \cite{sha16} \\ \hline
\end{tabular}
\end{center}

\end{tablehere}

If the number distribution of galaxies $\Phi(X)$ as a function of the
galactic property $X$ can be observationally determined, then the mass
function of MBHs can be estimated as
\be
\frac{\d N}{\d \log M_\bh}=\int \Phi(X) P(\log M_\bh|X) \d X, 
\ee
where $P(\log M_\bh|X)$, the probability distribution of $\log M_\bh$
at a given $X$, is assumed to be described by a Gaussian function,
i.e., 
\be
P(\log M_\bh|X)=\frac{1}{\sqrt{2\pi} \Delta_{\log M_\bh}} \exp\left[
-\frac{(\log M_\bh-\langle \log M_\bh \rangle)^2}{2\Delta_{\log
M_\bh}^2} \right],
\label{eq:m-p}
\ee
where $\left< \log M_\bh \right>$ is the mean value of $\log M_\bh$,
which can be determined by the $M_\bh -X$ relation obtained from
observations [see Equation~(\ref{eq:m-x})], and $\Delta_{\log M_\bh}$ 
is the intrinsic scatter of the relation.

\subsection{Distribution function of velocity dispersion of elliptical
galaxies and spheroids}

Sheth et al. \cite{she03} first estimate the distribution function of 
velocity dispersion (VDF) of early type galaxies in the local universe
according to a large sample ($\sim 10^4$) of early type galaxies
observed by the Sloan Digital Sky Survey (SDSS), and they found that
the VDF can be fitted by the Schechter function. They also
estimated the VDF for late type galaxies, according to the relations
between luminosity, circular velocity and velocity dispersion, and
the luminosity function of late type galaxies. With these estimates,
they obtained the VDF for all galaxies (all ellipticals and
spheroids). Adopting similar method as that in Sheth et al. \cite{she03},
Mitchell et al. \cite{mit05} and Choi et al. \cite{cho07} also estimated 
the VDF for early type galaxies by using even larger SDSS samples, and 
they also obtained the VDF for all galaxies by similarly considering the 
contribution from late type galaxies.  In a recent work, Bernardi et al. 
\cite{ber10} and Sohn et al. \cite{soh17} estimated the VDF
for all galaxies directly. As shown in Table~\ref{tab:t2}, all those
four estimates about the VDF for all galaxies are adopted to calculate
the MBH mass density in the local universe according to
Equations~(\ref{eq:mbhden})-(\ref{eq:m-p}).

\subsection{Stellar mass function of ellipticals and spheroids}

The stellar mass function (SMF) of nearby galaxies with different
morphologies have also been determined by SDSS observations
\cite{ber10}, and the fraction $f^i$ of $i-$th morphology at
a given mass can then be obtained. Thus the SMF for spheroids can be
derived if the bulge-to-total mass ratio ($B/T$) is given, i.e.,
\be
\phi(M_{*, \sph})=\sum_i f^i(M_{*, \rm tot}) \phi(M_{*, \rm tot})
\frac{\d M_{*, \rm tot}}{\d M_{*, \sph}}, 
\ee
where $M_{*, \sph} = (B/T) M_{*, \rm tot}$, and $\phi(M_{*, \rm tot})$
is the SMF for all galaxies (for a more detailed description about the
derivation of the SMF of ellipticals and spheroids, see Zhang et al.
\cite{zha12}). According to Weinzirl et al. \cite{wei09}, the 
bulge-to-total mass ratio $B/T$ for E, S0, Sa, Sb, and Scd are $1$, 
$0.28\pm0.02$, $0.46\pm0.05$, $0.22\pm0.08$, and $0.15\pm0.05$, 
respectively. Li and White \cite{li09} and Moustakas et al. \cite{ mou13} 
also estimated the SMF for all galaxies, and their SMF can also be used 
to estimate the SMF of ellipticals and spheroids by assuming the same 
fraction of each morphology at a given stellar mass as that of Bernardi 
et al. \cite{ber10}. In addition, Thanjavur et al. \cite{tha16} recently 
directly determined the SMF of ellipticals and spheroids in the local 
universe.  As shown in Table~\ref{tab:t3}, all those four estimates about 
the SMF for ellipticals and spheroids are also adopted to calculate the 
MBH mass density in the local universe according to 
Equations~(\ref{eq:mbhden})-(\ref{eq:m-p}).

Tables~\ref{tab:t2} and \ref{tab:t3} list all the estimated values for
the local MBH mass densities obtained through the method introduced above,
by combining various $M_\bh - \sigma$ and $M_\bh - M_{*, \sph}$
relations with the VDFs and SMFs, respectively. As seen from
Table~\ref{tab:t2}, the difference in the MBH mass density estimates
induced by adopting the $M_\bh - \sigma$ relation determined by
different authors can be up to a factor of $5-6$, which is mainly due
to the difference in the normalizations of the $M_\bh - \sigma$
relation obtained by different authors.  The difference in the MBH
mass density estimates induced by adopting different estimates of the
VDF is only about $20\%$ to $70\%$, which is significantly smaller
than that induced by the uncertainties in the determination of the
$M_\bh - \sigma$ relation.  As seen from Table~\ref{tab:t3}, the
difference in the MBH mass density estimates induced by adopting the
$M_\bh - M_{*,\rm sph}$ relation determined by different authors can
be as large as a factor of $15$, while the difference in the MBH mass
density estimates induced by adopting different SMFs determined by
different authors is no more than $80\%$. It is clear that the largest
uncertainty in the MBH mass density estimates is due to the
uncertainty in the determination of the relation between the masses of
MBHs and the properties of their host galaxies, especially the
normalization of these relations.

\begin{tablehere}
\caption{The MBH mass density in the local universe ($\rho_{\bh,0})$
inferred from the $M_\bh - \sigma$ relation and the
$\sigma-$distribution function. The second, to sixth columns show the
MBH mass densities in the local universe $\rho_{\bh,0}$, which are
estimated by adopting the $\sigma-$distribution function presented in
Sheth et al. \cite{she03}, Mitchell et al. \cite{mit05}, Choi et al. 
\cite{cho07}, Bernardi et al. \cite{ber10}, and Sohn et al. \cite{soh17},
respectively, as labeled in the first row. The second to eighth rows
show $\rho_{\bh,0}$ which are estimated by adopting the $M_\bh -
\sigma$ relation presented in Tremaine et al. \cite{tre02}, G\"{u}ltekin 
et al. \cite{gul09}, Graham et al. \cite{gra12}, McConnell and Ma 
\cite{mcc13}, Kormendy and Ho \cite{kor13}, Saglia et al. \cite{sag16}, 
and Shankar et al. \cite{sha16}, respectively, as listed in
Table~\ref{tab:t1}. The unit of $\rho_{\bh,0}$ is $10^5 M_\odot~{\rm
Mpc}^{-3}$. The errors are obtained by considering the $1-\sigma$
uncertainty both in the estimates of the $\sigma-$distribution and 
the $M_\bh - \sigma$ relation.}
\label{tab:t2}

\renewcommand{\arraystretch}{1.5}
\vspace{-1mm}\footnotesize
\centering
\begin{center} \doublerulesep 0.1pt \tabcolsep 6.5pt
\begin{tabular}{cccccc}
\hline 
	 & She03 	
& Mit05 	& Cho07		& Ber10    &  Soh17 \\   	    \hline
Tre02 	&$3.0^{+0.6}_{-0.5}$		& $2.3^{+0.5}_{-0.4}$	 &
$3.0^{+0.7}_{-0.5}$	& $3.6^{+0.7}_{-0.6}$ & $2.2^{+0.4}_{-0.3}$    \\ 
Gul09 	&$3.7^{+1.0}_{-0.7}$		& $2.9^{+0.8}_{-0.6}$	&
$3.7^{+1.0}_{-0.8}$	&$4.5^{+1.2}_{-1.0}$	& $2.9^{+0.7}_{-0.5}$ 	\\
Gra12	&$4.1^{+2.8}_{-1.2}$		& $3.7^{+1.3}_{-1.5}$	&
$5.3^{+1.6}_{-2.5}$ 	& $7.1^{+0.9}_{-3.7}$ 	& $3.4^{+1.3}_{-1.1}$	\\
McC13	&$4.9^{+0.8}_{-0.7}$		& $3.6^{+0.6}_{-0.5}$	&
$5.1^{+0.8}_{-0.7}$ 	& $6.4^{+1.0}_{-0.9}$ 	 & $4.3^{+0.7}_{-0.6}$	\\
KH13	&$6.5^{+1.0}_{-0.9}$		& $4.9^{+0.9}_{-0.7}$	&
$6.5^{+1.0}_{-0.9}$ 	& $7.8^{+1.3}_{-1.2}$ 	 & $5.0^{+0.8}_{-0.7}$	\\
Sag16	&$5.6^{+0.6}_{-0.5}$		& $4.2^{+0.5}_{-0.4}$	&
$5.6^{+0.7}_{-0.6}$ 	& $6.9^{+0.8}_{-0.7}$ 	 & $4.6^{+0.5}_{-0.4}$	\\
Sha16	&$1.3\pm0.1$			& $0.9\pm0.1$
& $1.3\pm0.2$ 			& $1.6\pm0.1$ 		& $1.1\pm0.1$ 	\\
\hline						
\end{tabular}
\end{center}

\end{tablehere}

\section{Total energy density radiated by AGNs/QSOs over the cosmic time}
\label{sec:agn}

According to the So{\l}tan argument \cite{sol82, yu02}, the
total energy density radiated by AGNs/QSOs ($E_{\rm tot}$) can be
obtained by integrating the AGN/QSO luminosity function over the
cosmic time. The AGN/QSO luminosity function determined in the hard
X-ray band (XLF) is more complete than that determined in the optical
band (OLF) because many AGNs/QSOs may be missed in the optical surveys
due to obscurations. Therefore, XLF is favored in estimating $E_{\rm
tot}$ via the So{\l}tan argument, i.e., 
\be
E_{\rm tot} = \int 
\left|\frac{\d t}{\d z}\right| 
\d z \int\int C_\x L_\x  P(C_\x|L_\x)  \phi(L_\x, z)  \d C_\x \d L_\x,  
\label{eq:Etot}
\ee
where $C_\x \equiv L_\bol/L_\x$ is the bolometric correction at the
X-ray band, $P(C_\x| L_\x)$ is the probability distribution of the
bolometric correction for AGNs/QSOs with a given $L_\x$, and
$\left|\frac{\d t}{\d z}\right|=H^{-1}_0 \sqrt{\Omega_{\rm M} (1+z)^3+
\Omega_\Lambda}$.

\begin{tablehere}
\caption{The MBH mass density in the local universe ($\rho_{\bh,0})$
inferred from the $M_\bh-M_{*, \sph}$ relation and the stellar mass
function (SMF). The second to fifth columns show the MBH mass densities 
in the local universe $\rho_{\bh,0}$, which are estimated by using the 
SMF of ellipticals and spheroids obtained based on the total SMF of 
Li and White \cite{li09}, Bernardi et al. \cite{ber10}, and Moustakas 
et al. \cite{mou13}, and that directly determined by Thanjavur et al. 
\cite{tha16}, respectively. The second to seventh rows show 
$\rho_{\bh,0}$, which are estimated by adopting the $M_\bh-M_{*, \sph}$ 
relation presented in H\"{a}ring and Rix \cite{har04}, Graham et al. 
\cite{gra12}, McConnell and Ma \cite{mcc13}, Kormendy and Ho \cite{kor13}, 
Saglia et al. \cite{sag16}, and Shankar et al. \cite{sha16} respectively, 
as listed in Table~\ref{tab:t1}. The unit of $\rho_{\bh,0}$ is $10^5 
M_\odot~{\rm Mpc}^{-3}$. The errors are obtained by considering the 
$1-\sigma$ uncertainty both in the estimates of the SMF and in the \
$M_\bh - M_{*,\rm sph}$ relation.} 
\label{tab:t3}
\renewcommand{\arraystretch}{1.5}
\vspace{-1mm}\footnotesize
\centering
\begin{center} \doublerulesep 0.1pt \tabcolsep 6.5pt
\begin{tabular}{ccccc}
\hline 
	 & LW09 		
& Ber10		& Mou13		& Tha16     \\   	    \hline
Har04 	&$2.2^{+0.6}_{-0.5}$		& $3.8^{+1.8}_{-0.9}$	&
$2.7^{+0.7}_{-0.6}$ 	&$3.6^{+1.0}_{-0.8}$		 	\\
%
Gra12	&$4.6^{+4.5}_{-1.9}$		& $8.7^{+12.2}_{-4.1}$	&
$6.1^{+6.8}_{-2.8}$ 	& $9.0^{+8.1}_{-3.4}$ 		\\
%
McC13	&$3.7^{+1.4}_{-1.0}$		& $7.5^{+4.9}_{-2.7}$	&
$5.0^{+1.8}_{-1.3}$ 	& $6.6^{+2.1}_{-1.7}$ 	 	\\
%
KH13	&$6.2^{+1.2}_{-0.9}$		& $11.2^{+4.6}_{-2.4}$	&
$7.9^{+1.4}_{-1.2}$ 	& $10.5^{+1.7}_{-1.5}$ 	 	\\
%
Sag16	&$8.1^{+1.3}_{-1.0}$		& $12.0^{+3.6}_{-1.7}$	&
$9.3^{+1.3}_{-1.0}$ 	& $11.9^{+1.1}_{-0.9}$ 	 	\\
%
Sha16	&$0.55\pm0.1$			& $1.4^{+0.6}_{-0.3}$	&
$0.84\pm0.1$ 		& $1.0\pm0.1$ 			 	\\
\hline			
\end{tabular}
\end{center}

\end{tablehere}

\begin{tablehere}
\caption{Estimates of the total energy radiated from AGNs/QSOs across
the cosmic time. The total energy density $E_{\rm tot}/c^2$ is in
unit of $ 10^4 M_\odot \ \rm{Mpc^{-3}}$. The first column shows the values
of $E_{\rm tot}/c^2$, which are estimated by adopting the XLFs obtained 
in the references listed in the second column.}
\label{tab:t4}
\vspace{-1mm}\footnotesize
\begin{center} \doublerulesep 0.1pt \tabcolsep 19.5pt
\begin{tabular}{ccc}
\hline 

$E_{\rm tot}/c^2$	 		& References
\\   	    \hline
$5.2\pm0.7$		& Ueda+03 \cite{ued03}		\\
$5.6\pm0.6$		& La Franca+05  \cite{laf05} \\
$3.4\pm0.2$		& Silverman+08 \cite{sil08} \\
$4.6\pm0.4$		& Ebrero+09 \cite{ebr09} \\
$3.0\pm0.5$		& Yencho+09 \cite{yen09} \\
$3.7\pm0.6$		& Aird+10 \cite{air10}	\\
$6.1\pm1.4$		& Aird+15 \cite{air15} \\
$5.9\pm2.3$		& Fotopoulou+16 \cite{fot16} \\
$5.4\pm1.2$		& Ranalli+16 	\cite{ran16}	\\ \hline		
\end{tabular}
\end{center}

\end{tablehere}

The XLFs of AGNs/QSOs have been frequently measured/estimated in a
large redshift range according to observations by different surveys 
\cite{ued03, laf05, sil08, ebr09, yen09, air10, air15, fot16, ran16}. 
Most XLFs given in the literature are for the $2-10$ keV band, but 
some are for a slightly different band, e.g., $2-7$ keV,  $2-8$ keV, 
or $5-10$ keV band. We adopt the bolometric corrections for the $2-10$ 
keV band as a function of   $L_\x$ given by Zhang et al. \cite{zha12}, 
which is obtained by using similar procedures as those in Hopkins et al. 
\cite{hop07} when constructing the template spectrum, although it is 
fitted as a function of $L_\x$ rather than $L_{\rm bol}$.  For those 
XLFs in other X-ray bands, we convert it to the $2-10$ keV band by 
assuming a canonical photon index $\Gamma =1.8$. We extrapolate those 
XLFs to high redshifts and low luminosities and obtain $E_{\rm tot}$ 
according to Equation~(\ref{eq:Etot}), as listed in Table~\ref{tab:t4}. 
As seen from Table~\ref{tab:t4}, the uncertainty in the $E_{\rm tot}$ 
estimates can be as large as a factor of $2$ due to the uncertainties 
in the determinations of the XLFs.

\section{Mean radiative efficiency of AGNs/QSOs}
\label{sec:result}

It is generally believed that MBHs in galactic centers obtain their
masses mainly through the accretion in the bright AGN/QSO phases
\cite{yu02, mar04}. Assuming that the initial masses of the
seeds of MBHs are much smaller than their final masses, then the mean
radiative efficiency of AGNs/QSOs is given by
\be
\epsilon \simeq \frac{1}{1+\rho_{\bh,0} c^2 /E_{\rm tot} },
\label{eq:eff}
\ee
where $c$ is the speed of light. According to $\rho_{\bh,0}$ and 
$E_{\rm tot}/c^2$ listed in Table~\ref{tab:t2}, \ref{tab:t3}, and
\ref{tab:t4}, respectively, we can obtain the estimates of $\epsilon$ 
according to Equation~(\ref{eq:eff}). Since there are seven different 
determinations of the $M_\bh-\sigma$ relations, five different 
determinations of the VDFs, and nine estimates of the total density 
of the energy radiated from AGNs, we have $7\times5\times 9= 315$ 
estimates of $\epsilon$ based on the $M_\bh-\sigma$ relation and
the VDFs; and since there are six different determinations of the 
$M_\bh-M_{*, \sph}$ relations and four different determinations of 
the SMFs, we have $6 \times 4 \times 9= 216$ estimates of $\epsilon$ 
based on the $M_\bh-M_{*, \sph}$ relation and the SMFs. All those
estimates for the mean efficiency are shown as colored circles in 
the top and bottom panels of Figure 1.

\begin{figure}[H]
\centering
\includegraphics[scale=0.68]{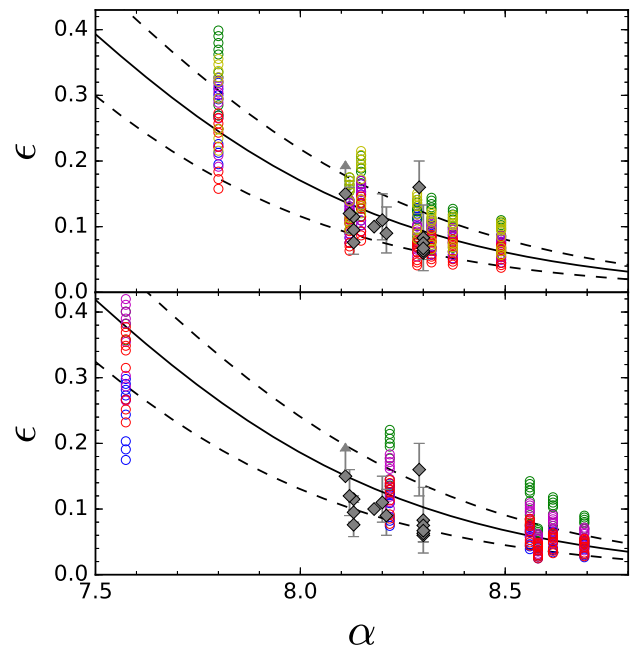}
\caption{Mean radiative efficiency on the $\epsilon - \alpha$ plane
($\alpha$ is the normalization of the relation between the masses of
MBHs and the properties of their host galaxies). Top and bottom panels
show the estimates of $\epsilon$ by using the $M_\bh - \sigma$ relation 
and the $M_\bh - M_{*,\rm sph}$ relation, respectively. In top panel, 
the blue, green, magenta, red and yellow open circles represent 
$\epsilon$ estimated by using the VDFs obtained in Sheth et al.
\cite{she03}, Mitchell et al. \cite{mit05}, Choi et al. \cite{cho07},
 Bernardi et al. \cite{ber10} and Sohn et al. \cite{soh17}, respectively; 
while in the bottom panel, the green, blue, magenta, and red open circles
represent $\epsilon$ estimated by using the SMFs obtained in Li and White 
\cite{li09}, Bernardi et al. \cite{ber10}, Moustakas et al. \cite{mou13}, 
and Thanjavur et al. \cite{tha16}, respectively. The solid line in the top 
panel shows the mean result obtained by setting the slope of the relation 
$\beta \sim 3.7 - 5.7$ and using the VDF of Bernardi et al. \cite{ber10} 
and XLF of Ranalli et al. \cite{ran16}; while the solid line in the bottom 
panel represents the mean results obtained by setting the slope of the 
relation $\beta \sim 0.75 -1.95$, and using  the using latest SMF 
\cite{tha16} and XLF \cite{ran16}.  The dashed lines in both panels show 
the $1-\sigma$ uncertainty of the estimates by considering the errors in 
both the VDF/SMF and XLF.  For comparison, the values of mean efficiency
obtained in the literature are also shown as grey diamonds.} 
\label{fig:eff}
\end{figure}

As discussed in section~\ref{sec:bhmf}, the largest uncertainty in the 
estimate of the MBH mass density in the local universe is caused by 
the uncertainty in the determination of the normalization of the 
relation between the the masses of MBHs ($M_\bh$) and the properties 
of their host galaxies ($X$).  Figure~\ref{fig:eff} shows the estimated 
$\epsilon$ on the $\epsilon$ vs. $\alpha$ plane ($\alpha$ is the 
normalization of the $M_\bh - X$ relation). As seen from Figure~\ref{fig:eff}, 
the estimated values of $\epsilon$ cover a wide range, from $0.03$ to
$\sim 0.40$, due to the application of different versions of the
$M_\bh - X$ relations, VDFs, SMFs, and XLFs. {\it However, we note
here that the range of the estimated $\epsilon$ is fully consistent
with theoretical expected values of the efficiency for thin disk
accretion onto MBHs with spins in the range from $0$ to $1$.} For
fixed VDFs/SMFs and XLFs, the estimated values of $\epsilon$ can
differ by a factor of $\sim 6$ if adopting different versions of the
$M_\bh -X$ relation; while for a fixed $M_\bh - X$ relation, the
estimated values of $\epsilon$ can differ by a factor at most $\sim 3$
if adopting different versions of the VDFs/SMFs and XLFs of AGNs. 

To investigate how the efficiency depends on the normalization of
$M_\bh-M_{*, \sph}$ relation or the $M_\bh - \sigma$ relation, we
adopt one of the VDFs/SMFs and AGN/QSO XLFs and arbitrarily set an
$\alpha$ and let $\beta$ varying in the range from $ 3.7$ to $5.7$ for
$M_\bh-\sigma$ and from $ 0.75$ to $1.95$ for  the $M_\bh-M_{*, \sph}$
relation, and then we get the estimates of $\epsilon$.  The solid line
and the dashed lines in Figure~\ref{fig:eff} show the mean and the
$1-\sigma$ uncertainty of the estimates with the considerations of
errors in both the  VDF/SMF and XLF. These lines indicate that the
uncertainty of the normalizations of the $M_\bh -X$ relation dominates
the errors in the estimates of the mean radiative efficiency.

As discussed above, the uncertainties in the estimates of the mean
radiative efficiency ($\epsilon$) of AGNs/QSOs are quite large, which
is caused by various reasons, e.g., the uncertainties in the
determinations of the relation between the MBH masses and the
properties of the MBH host galaxies, in the determinations of the
distribution function of galactic properties, and in the estimates of
the AGN/QSO luminosity functions. In order to get a more accurate
estimate of the mean radiative efficiency, we need to do some
statistical analysis. Assuming that all the estimates of $\epsilon$
above are: (1) independent and (2) without systematic effects. These
two assumptions are very likely to be valid as the estimates of
$\epsilon$ are obtained from a  number of independent measurements of
the galaxy and AGN/QSO properties, and independent determinations of the
relations between the masses of MBHs and the properties of the MBH
host galaxies. It is naturally expected that half of the estimates are
below the true value and the other half are above the true value.
Therefore, we assume the median value obtained from the large
number of $\epsilon$ estimates in this paper is close to the true
value of $\epsilon$ (and it is the true value if the number of estimates
is infinity). This type of median statistics is a powerful tool to
deal with a large number of measurements with substantial
uncertainties, which has been demonstrated to be extremely useful by
Gott et al. \cite{got01}, e.g., in analyzing the
measurements of the Hubble constant and others.

\begin{figure}[H]
\centering
\includegraphics[scale=0.68]{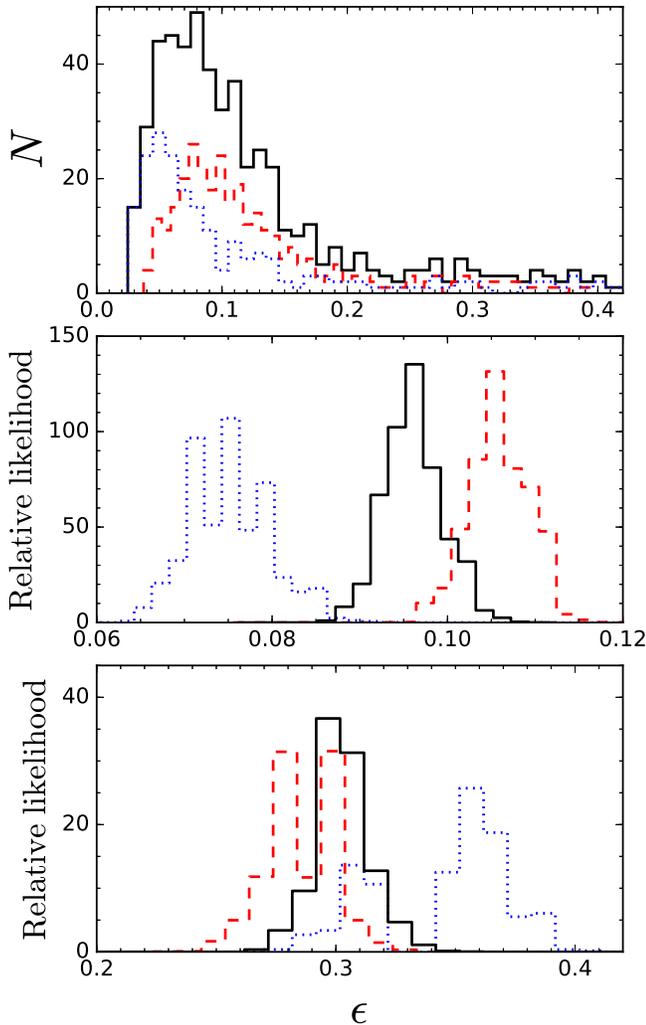}
\caption{Number distribution of the estimated values for the mean
radiative efficiency $\epsilon$ (top panel) and the likelihood
distribution of the median of $\epsilon$ (middle panel). The red
dashed histogram is for those $\epsilon$ estimated by using the
$M_\bh-\sigma$ relation, with median efficiency
$0.105^{+0.006}_{-0.008}$.  the blue dotted is for those $\epsilon$
estimated by using the $M_\bh-M_{*, \sph}$ relation, with median value
$0.074^{+0.010}_{-0.008}$. The black solid is for all the estimates,
and the median efficiency is $0.095^{+0.006}_{-0.007}$. Bottom panel:
likelihood distribution of the median efficiency which are derived by
applying only the $M_\bh-\sigma/M_{*, \sph}$ relation of Shankar et al.
\cite{sha16}. Line styles are similar to the top two panels. The
median efficiency is $0.272^{+0.032}_{-0.024}$ and
$0.349^{+0.033}_{-0.062}$ for the red dashed and blue dotted lines,
and is $0.295^{+0.025}_{-0.022}$ for the black solid line. All of the
errors above are given at 95\% confidence level.} 
\label{fig:ne}
\end{figure}

For those estimated $\epsilon$, we rank them as $\epsilon_i$ by value
such that $\epsilon_j > \epsilon_i$ if $j>i$. Then the true value of
the median lies in between $\epsilon_i$ and $\epsilon_{i+1}$ with a
probability of
\be
P = \frac{2^{-N} N!}{i! (N-i)!}
\label{eq:prob}
\ee According to Equation~(\ref{eq:prob}), we obtain the true
probability distribution of the median of the estimated $\epsilon$.

Figure~\ref{fig:ne} shows both the number distribution of the
estimated value of the mean radiative efficiency $\epsilon$ (top
panel) and the probability distributions of the median value of all
the estimated $\epsilon$ (middle panel).  The red dashed, blue dotted, 
and black solid histograms in both the two panels show results for
$\epsilon$ estimated by using the $M_\bh - \sigma$ relation, the
$M_\bh - M_{*,\rm sph}$ relation, and all estimated $\epsilon$,
respectively.  As seen from Figure~\ref{fig:ne}, the probability
distributions of the median $\epsilon$ in the middle panel are much
narrower than the number distributions of $\epsilon$ shown in the top
panel.  According to the probability distributions of the median value
of all the estimated $\epsilon$, we find that the median $\epsilon$ is
$0.095^{+0.006}_{-0.007}$, where the errors mark the $95\%$
confidence level of the median. If only adopting the values of
$\epsilon$ estimated by using the $M_\bh - \sigma$ relation or the
$M_\bh - M_{*,\rm sph}$ relation, the median value is then
$0.105^{+0.006}_{-0.008}$ or $0.074^{+0.010}_{-0.008}$, respectively.
The median obtained by using the $M_\bh - M_{*,\rm sph}$ relation is
substantially smaller than that by the $M_\bh - \sigma$ relation.
Note that Yu and Tremaine \cite{yu02} first pointed out that there 
could be some biases in the determined relation between $M_\bh$ and 
$ M_{*,\rm sph}$, which is lately confirmed by Tundo et al.  
\cite{tun07}. This bias may lead to an overestimation of the total 
MBH mass density by using the $M_\bh - M_{*,\rm sph}$ relation. If 
only considering the estimates of $\epsilon$ from the $M_\bh - \sigma$ 
relation, then $\epsilon =0.105^{+0.006}_{-0.008}$, consistent with 
the canonical value $0.1$ \cite{yu02}.

The above results obtained from the median statistics are valid if
there are no systematic biases in all of the measurements of the
$M_\bh-\sigma/M_{*, \sph}$ relations, VDFs/SMFs, and XLFs, which is
probably true as those are generally determined by different authors
and with (at least partly) different data. One exception is the
$M_\bh-\sigma/M_{*, \sph}$ relation discussed in Shankar et al. 
\cite{sha16}. They found that the nearby galaxy sample with MBH mass
measurements could be biased from that of the SDSS samples, and the
SDSS galaxies have significant higher $\sigma$ than local galaxies
with similar stellar mass, as also noted earlier by Yu and Tremaine 
\cite{yu02} and Tundo et al. \cite{tun07}. We therefore repeat the 
statistical analysis but only for the efficiencies inferred from 
$M_\bh-\sigma/M_{*, \sph}$ relation of Shankar et al. \cite{sha16}. 
The likelihood distribution is shown in the bottom panel of 
Figure~\ref{fig:ne}. The median efficiency is $0.272^{+0.032}_{-0.024}$ 
and $0.349^{+0.033}_{-0.062}$ when the $M_\bh - \sigma$ relation and 
the $M_\bh - M_{*,\rm sph}$ relation respectively is solely used, 
and for the total of them, the median value is $0.295^{+0.025}_{-0.022}$. 

Note that a fraction of Compton-thick AGNs/QSOs may be still missed
from the surveys in the hard X-ray band ($\sim 2-10$\,keV)
\cite{mal09}.  The fraction of the missed Compton-thick
AGNs/QSOs could  be $\sim 20\%$ of the total AGN/QSO population
\cite{mal09}. If this is true, then the total energy radiated from
AGNs/QSOs over the cosmic time ($E_{\rm tot}$) may be underestimated
by $20\%$ in the above analysis, which would lead to an
underestimation of the mean radiative efficiency by $\sim 20\%$. The
inclusion of those Compton-thick objects may increase the estimated
value of the mean radiative efficiency to $\sim 0.12$.

According to the mean radiative efficiency of AGNs/QSOs estimated in
this paper ($\epsilon = 0.105$ or $\sim0.12$ by assuming a fraction of 
$\sim 20\%$ Compton-thick objects), which corresponds to an effective 
spin of those MBHs $a \sim 0.71$ (or 0.80) if assuming all MBHs in
AGNs/QSOs accreting material via the standard thin disk. If only adopting 
those estimates of $\epsilon$ by using the $M_\bh - \sigma/M_{*,\rm sph}$ 
relation of Shankar et al. \cite{sha16}, then mean radiative efficiency 
is $\sim 0.295$, which would correspond to an effective spin of those 
MBHs $a \gtrsim 0.99$ by further considering the contribution from the 
missed Compton-thick AGNs/QSOs.

Currently there are about two dozens of MBHs in low redshift AGNs/QSOs
that have spin measurements according to the Fe K$\alpha$ line
detections \cite{bre13, rey14}. Those MBHs all have
intermediate to extremely high spins, i.e., $a \sim 0.4-1$.  Those
measurements may be still not conclusive, but they seem to be
consistent with our estimates about the mean radiative efficiency for
the whole population of AGNs/QSOs in this paper. We also note that the
radiative efficiency of individual SDSS QSOs has also been estimated
by Wu et al. \cite{wu13}, in which the efficiency is estimated by using 
the bolometric luminosity estimated from multi-wavelength spectra and 
the accretion rate estimated by fitting the continuum spectra to
specific disk accretion model.  Wu et al. \cite{wu13} obtained the mean
efficiency of the SDSS QSOs to be about $0.11 - 0.16$, which is also
consistent with the global constraint on the mean radiative efficiency
of whole population of AGNs/QSOs obtained in the present paper.

\section{Conclusions}
\label{sec:conclusion}

The mean radiative efficiency of accretion onto massive black holes
(MBHs) in AGNs and QSOs can be estimated by using the So{\l}tan
argument, by matching the mass density of MBHs in the local universe 
to the accreted mass density by MBHs during AGN/QSO phases. In this
paper, we estimate the local MBH mass density through a combination of
various determinations of the correlations between the masses of MBHs
and the properties of MBH host galaxies, with the distribution functions 
of those galaxy properties. We also estimate the total energy density 
radiated by AGNs and QSOs by using various AGN/QSO X-ray luminosity 
functions in the literature. We obtain several hundred estimates of 
the mean radiative efficiency of AGNs/QSOs. 
Under the assumption that all those estimates are independent
of each other and free of systematic effects, we apply the median
statistics as described by Gott et al.\cite{got01} and find the mean radiative
efficiency of AGNs/QSOs is  $0.095^{+0.006}_{-0.007}$. Considering that
roughly $20\%$ Compton-thick objects may be missed from current
available X-ray surveys, the mean radiative efficiency may be as high as
$\sim 0.11$. However, the efficiency estimated by using
the $M_\bh - M_{*,\rm sph}$ relation may be biased. If only those
estimates of the efficiency by using the $M_\bh - \sigma$ relation
are used, then we find the mean radiative
efficiency of AGNs/QSOs is  $0.105^{+0.006}_{-0.008}$, which is
consistent with the canonical value $\sim0.1$. Considering the correction
due to those Compton-thick objects, the true mean radiative efficiency
may be $\sim 0.12$.

\vspace*{2mm} \Acknowledgements{\bahao This work is partly supported 
by the National Key Program for Science and Technology Research and 
Development (Grant No. 2016YFA0400704), the National Natural Science 
Foundation of China under grant No.11373031 and 11390372, and the 
Strategic Priority Program of the Chinese Academy of Sciences (Grant 
No. XDB 23040100).}

\end{multicols}

\end{document}